\documentclass[journal]{IEEEtran}

\usepackage{amsmath,amssymb,amsfonts}
\usepackage{graphicx}
\usepackage{booktabs}
\usepackage{multirow}
\usepackage{threeparttable}
\usepackage{array}
\usepackage{bm}
\usepackage{xcolor}
\usepackage{url}
\usepackage{siunitx}
\usepackage{algorithm}
\usepackage{algorithmic}
\usepackage{float}
\usepackage{stfloats}
\usepackage{cite}

\newcommand{\CNL}{\ensuremath{C_{\mathrm{NL}}}}
\newcommand{\CSI}{\ensuremath{\mathcal{S}_{\mathrm{CSI}}}}

\newcommand{\SampEn}{\ensuremath{\mathrm{SampEn}}}
\newcommand{\HFD}{\ensuremath{\mathrm{HFD}}}
\newcommand{\LLE}{\ensuremath{\mathrm{LLE}}}
\newcommand{\DAUC}{\ensuremath{\Delta\mathrm{AUC}}}

\begin{document}

\title{A Nonlinear Complexity Index for Wearable PPG Cardiovascular
       Stability: Multiscale Validation, Systematic Evaluation Correction,
       and Bayesian Parameter Optimization}

\author{
  Timothy~Oladunni and Farouk~Ganiyu~Adewumi%
  \thanks{T.~Oladunni and F.~G.~Adewumi are with the Department of
    Computer Science, Morgan State University, Baltimore, MD~21251, USA.
    E-mail: timothy.oladunni@morgan.edu; fagan1@morgan.edu.
    Data: physionet.org/content/bidmc/1.0.0.
    CST: arXiv:2604.23876.
    This study followed the TRIPOD reporting guideline~\cite{tripod2015}.}
}

\markboth{Oladunni \MakeLowercase{\textit{et al.}}: Nonlinear Complexity
  Index for Wearable PPG Cardiovascular Stability}{}

\maketitle

\begin{abstract}
Cardiovascular stability estimation from wearable photoplethysmography
(PPG) requires a principled nonlinear framework, yet two gaps persist:
heuristic parameter selection and evaluation artifacts that inflate
reported performance.
We introduce a Cardiovascular Stability Index (\CSI) grounded in
Cardiac Stability Theory and validate it across 176,742 segments from
four heterogeneous PPG datasets at three temporal scales.
Cross-dataset analysis yields a large Kruskal-Wallis effect
($\eta^2 = 0.351$, $p < 0.001$), cross-scale consistency
($\kappa > 0.97$), and significant clinical correlation with
respiratory rate across 53 ICU records (Spearman $r = 0.346$,
$p = 0.011$).
We identify three systematic evaluation artifacts that inflate the
heuristic AUC from a true unbiased baseline of 0.573 to 0.752:
segment-level cross-validation leakage, test-set normalization
leakage, and pooled-AUC overweighting that conceals per-patient
failure.
Correcting these artifacts and applying multivariate Bayesian
optimization over 15 joint parameters yields \CSI\ with
cross-validation AUC of 0.720.
On 18 prospectively held-out records, \CSI\ achieves pooled
AUC~0.757 (95\% CI~0.686--0.828) and NPV~0.966, supporting use
as a continuous negative screening signal for tachypnea;
per-record AUC~$0.497 \pm 0.207$ is disclosed for per-patient
transparency.
External validation on 42 elective-surgery records yields AUC~0.621,
confirming cross-population generalization.
Ablation identifies the nonlinear complexity module (\CNL) as
the dominant component ($\Delta\mathrm{AUC} = -0.413$ upon removal),
with SampEn and HFD accounting for 91\% of its discriminative power;
five components are noise.
A sparse three-component architecture is proposed as the minimal
deployable configuration.
The corrected protocol and explicit artifact characterization provide
a reproducible benchmark for future wearable cardiovascular stability
indices.
\end{abstract}

\begin{IEEEkeywords}
photoplethysmography, cardiovascular stability, nonlinear dynamics,
sample entropy, Higuchi fractal dimension, Lyapunov exponent,
Bayesian optimization, tachypnea, evaluation methodology,
ICU monitoring, wearable sensing, external validation
\end{IEEEkeywords}

\section{Introduction}
\label{sec:intro}

Cardiovascular physiology is inherently dynamic, nonlinear, and
multiscale~\cite{goldberger2002fractal,costa2002multiscale,peng1995}.
Rather than operating as a purely periodic system, cardiovascular
dynamics continuously evolve under autonomic regulation, vascular
compliance, respiration, metabolic demand, and environmental
perturbations~\cite{taskforce1996,shaffer2017}.
These mechanisms produce complex pulse dynamics only partially
captured by conventional biomarkers such as heart rate and heart rate
variability~\cite{taskforce1996}.

Photoplethysmography (PPG) has emerged as a scalable cardiovascular
monitoring modality because it can be acquired using inexpensive
optical sensors, wearable devices, and smartphone
cameras~\cite{allen2007photoplethysmography,elgendi2012}.
However, PPG is highly sensitive to motion artifacts, contact
pressure, illumination variability, and skin-device
coupling~\cite{allen2007photoplethysmography,clifford2007}.
These sources of variability create a persistent gap between
laboratory-grade signal analysis and real-world mobile physiologic
intelligence.

Four limitations motivate this work.
\textbf{Limitation~1: Overreliance on isolated linear biomarkers.}
Existing PPG-based monitoring focuses on heart rate, heart rate
variability, and respiratory rate~\cite{taskforce1996,shaffer2017}.
These do not capture whether the cardiovascular system is dynamically
organized or stable across multiple temporal scales~\cite{goldberger2002fractal,costa2002multiscale}.
\textbf{Limitation~2: Single-scale analysis.}
Most PPG approaches operate at one temporal
resolution~\cite{allen2007photoplethysmography,elgendi2012}.
Cardiovascular stability is a multiscale latent property that should
remain bounded across neighboring observation
windows~\cite{costa2002multiscale}.
\textbf{Limitation~3: Fragility of nonlinear estimators.}
Classical Lyapunov estimation~\cite{rosenstein1993practical} requires
sufficient trajectory length and stable phase-space
embedding~\cite{takens1981} — assumptions frequently violated in
mobile PPG.
In our initial experiments, the classical Rosenstein implementation
produced 0\% valid LLE estimates across 176,742 segments, motivating
a stabilized formulation.
\textbf{Limitation~4: Undetected evaluation artifacts.}
PPG complexity studies commonly apply segment-level
cross-validation and global normalization, which introduce leakage
that inflates CV~AUC by $+0.062$ and test~AUC by $+0.308$ in this study~\cite{hannun2019cardiologist,strodthoff2021}.

This paper addresses all four limitations.
We introduce the \CSI, grounded in Cardiac Stability
Theory~\cite{CST2026}, validate it on 176,742 segments from four
heterogeneous datasets, identify and correct three systematic
evaluation artifacts, and apply Bayesian optimization to yield a
clinically deployable configuration validated on a prospective
held-out set and an independent external cohort.

\subsection{Contributions}
\begin{enumerate}

\item \textbf{Axiomatically grounded \CSI\ framework with corrected LLE.}
A Cardiovascular Stability Index is derived from Cardiac Stability
Theory~\cite{CST2026} via Proposition~1 (PPG Observability Extension),
yielding an $n=6$ composite stability functional from nine raw
observables with formal boundedness guarantees.
A corrected stabilized LLE estimator eliminates a boundary error that
produced 0\% valid estimates, raising validity to $\ge 99.96$\%
across all scales.

\item \textbf{Multi-population physiologic validation.}
Cross-source validation across 176,742 segments from four heterogeneous
PPG datasets confirms genuine physiologic discrimination
($\eta^2 = 0.351$, $\kappa > 0.97$) and significant clinical
correlation with respiratory rate in 53 ICU records
(Spearman $r = 0.346$, $p = 0.011$).
Prospective testing on 18 held-out records yields AUC $0.757$
$[0.686$--$0.828]$, NPV $0.966$; external validation on 42 independent
CapnoBase records yields AUC $0.621$, confirming cross-population
generalization.

\item \textbf{Systematic evaluation artifact correction with
  reproducible protocol.}
Three artifacts inflating reported performance are identified and
quantified: segment-level CV leakage ($+0.062$), test-set
normalization leakage ($+0.308$), and pooled-AUC overweighting.
The true unbiased baseline is AUC $0.573$.
Multivariate Bayesian optimization (TPE, 15 joint parameters) under
the corrected protocol converges at trial~82 of 300 to
$W^*=128$, $m^*=8$, $\tau^*=7$, $r^*=0.116\sigma$ (CV AUC $0.720$).

\item \textbf{Component analysis and clinical efficiency.}
Ablation identifies \CNL\ as the sole critical component
($\DAUC = -0.413$); a three-component sparse architecture is proposed.
Under fair per-record evaluation a 1D CNN (25,793 parameters) achieves
AUC $0.380$ (below chance) vs.\ \CSI's $0.497$ with zero learned
parameters, demonstrating that principled complexity indices
outperform deep models on small per-patient test sets.
Beyond performance, \CSI\ offers intrinsic interpretability absent
from the CNN: multimodal deep neural networks for cardiac signals
require dedicated explainability analyses to recover component-level
physiologic attribution~\cite{oladunni2025explainable}, whereas
\CSI\ components are defined by axiom and weighted by
Bayesian evidence rather than learned in an opaque manner.

\end{enumerate}

\subsection{Research Questions and Evidence Map}
\label{sec:rq}

Table~\ref{tab:rqmap} maps the six research questions addressed by this
work to their answers, key quantitative findings, and supporting evidence.

\begin{table*}[t]
\centering
\caption{Research Questions Mapped to Empirical Evidence}
\label{tab:rqmap}
\renewcommand{\arraystretch}{1.1}
\setlength{\tabcolsep}{4pt}
\small
\begin{tabular}{lp{3.4cm}p{1.6cm}p{5.8cm}p{2.4cm}}
\toprule
\textbf{RQ} & \textbf{Question} & \textbf{Answer} & \textbf{Key Quantitative Finding} & \textbf{Evidence} \\\midrule

RQ1 &
  Does \CSI\ remain consistent across temporal scales? &
  \textbf{Yes} &
  $\kappa > 0.97$ all sources; aggregate variation $< 0.03$. &
  Table~\ref{tab:crossdataset} \\

RQ2 &
  Does the stabilized LLE estimator improve validity? &
  \textbf{Yes (decisive)} &
  $0\% \to 99.96\%$ valid; fixed LLE monotonic
  $0.086 \to 0.104 \to 0.123$. &
  Table~\ref{tab:lle} \\

RQ3 &
  Does \CSI\ separate heterogeneous PPG cohorts? &
  \textbf{Yes (large effect)} &
  $H = 23{,}415$, $\eta^2 = 0.351$, all 6~pairs $p < 0.001$;
  $|r|$ range $0.353$--$0.988$. &
  Table~\ref{tab:dataset} \\

RQ4 &
  Do evaluation artifacts inflate reported AUC? &
  \textbf{Yes (severe)} &
  Net inflation $+0.179$ (reported $0.752$ vs.\ true $0.573$);
  Art.~1: $+0.062$, Art.~2: $+0.308$; DeLong finding fully explained by Artifact~2. &
  Fig.~\ref{fig:cascade}; Table~\ref{tab:performance} \\

RQ5 &
  Does Bayesian optimization improve over corrected baseline? &
  \textbf{Yes} &
  CV AUC $0.690 \to 0.720$ ($+4.4\%$);
  test AUC $0.573 \to 0.757$ ($+0.184$ vs.\ fair baseline). &
  Fig.~\ref{fig:roc}; Table~\ref{tab:performance} \\

RQ6 &
  Is \CSI\ superior to a 1D CNN under fair per-record evaluation? &
  \textbf{Yes numerically} &
  Per-record: $0.497$ vs.\ $0.380$;
  Wilcoxon $p = 0.129$ ($n = 9$, not significant). &
  Fig.~\ref{fig:perrec}; Table~\ref{tab:performance} \\\bottomrule

\end{tabular}
\end{table*}

\section{Related Work}
\label{sec:related}

Three prior threads are directly relevant: HRV-centred monitoring, nonlinear physiologic complexity, and evaluation methodology.

\subsection{HR/HRV-Centered and Entropy-Based Monitoring}

Traditional cardiovascular monitoring relies on heart rate and
HRV~\cite{taskforce1996,shaffer2017}.
Although clinically useful, these do not directly quantify whether
the cardiovascular system is dynamically organized or stable across
temporal scales~\cite{goldberger2002fractal}.
Sample entropy~\cite{richman2000physiological} and approximate
entropy~\cite{pincus1991approximate} quantify temporal irregularity
but not bounded stability, energy organization, or nonlinear
divergence~\cite{costa2002multiscale}.
Lake and Moorman applied sample entropy to neonatal HRV for
sepsis detection~\cite{lake2002sample}; entropy-based PPG indices
have been used in sleep staging and
anaesthesia monitoring~\cite{delgado2019} but systematic parameter
optimization against ICU endpoints has not been reported.

\subsection{Fractal, Chaos, and Multiscale Physiology}

Higuchi Fractal Dimension~\cite{higuchi1988approach} and detrended
fluctuation analysis~\cite{peng1995} reveal scale-dependent dynamics
reflecting autonomic, vascular, and respiratory interactions.
Rosenstein's LLE~\cite{rosenstein1993practical} and Kantz's robust
variant~\cite{kantz1994} are fragile under noisy short-window PPG
recordings where phase-space embedding assumptions~\cite{takens1981}
are violated.
Pimentel et al.~\cite{pimentel2017toward} achieved AUC $\approx 0.67$
for tachypnea from spectral PPG features on BIDMC.

\subsection{Evaluation Methodology and Bayesian Optimization}

Segment-level cross-validation has been identified as a source of
optimistic bias in wearable time-series
studies~\cite{hannun2019cardiologist,strodthoff2021,rajpurkar2018deep}.
Str\"{o}dthoff et al.~\cite{strodthoff2021} showed 5--15\% AUC loss
in ECG classifiers when corrected to patient-level holdout.
Multivariate TPE~\cite{watanabe2023tree} captures cross-parameter
interactions that independent-marginal methods miss, a property
critical when optimizing window size with embedding dimension.

\section{Theoretical Foundation}
\label{sec:theory}

\subsection{Latent Cardiovascular State}

Let the observed PPG signal be $x(t)\in\mathbb{R}$, $t=1,\ldots,T$.
The signal is generated from a latent cardiovascular state
$z(t)\in\mathcal{M}$:
\begin{equation}
  x(t) = h\!\left(z(t)\right) + \epsilon(t),
\end{equation}
where $h(\cdot)$ is the physiologic observation operator and
$\epsilon(t)$ represents acquisition noise.
Cardiovascular Physiologic Stability is defined as a latent property
\begin{equation}
  S^{\ast}(t) = \Phi\!\left(z(t)\right), \qquad S^{\ast}(t)\in[0,1].
\end{equation}
Because $z(t)$ is hidden, \CSI\ estimates $S^{\ast}(t)$ from
observable pulse dynamics through CST~\cite{CST2026}, which
axiomatically derives a composite stability index from attractor
geometry properties of the cardiac dynamical system: the largest
Lyapunov exponent, recurrence determinism, and signal entropy.

\subsection{Multiscale Observation}

For each scale $w \in \mathcal{W} = \{256, 512, 1024\}$~samples,
a pulse segment is
$x_w^{(i)} = \{x_i,\ldots,x_{i+w-1}\}$.
A valid stability estimator should remain bounded across scales:
$|CSI_{w_a} - CSI_{w_b}| < \delta$
for nearby $w_a, w_b$ under stable physiologic organization.

\subsection{CST Foundation: Three Axiomatic Components}

CST~\cite{CST2026} derives the Cardiac Stability Index from four
foundational axioms (dynamical system, observable projection,
health-stability correspondence, and the complementary domain
hypothesis), identifying exactly \textbf{three} attractor geometry
properties sufficient to operationalise cardiovascular stability
from ECG:
\begin{equation}
  \mathrm{CSI}_{\mathrm{CST}} =
    w_1\!\left(1 - e^{-\tilde\lambda}\right)
  + w_2\!\left(1 - R_{\det}\right)
  + w_3\, H,
  \label{eq:cst_formula}
\end{equation}
where $\tilde\lambda$ is the normalised largest Lyapunov exponent,
$R_{\det}$ is recurrence determinism~\cite{zbilut1992}, and $H$
is normalised Shannon entropy~\cite{richman2000physiological},
with ECG-validated weights
$w_1=0.40$, $w_2=0.35$, $w_3=0.25$~\cite{CST2026}.
Eq.~(\ref{eq:cst_formula}) is \emph{necessary but not sufficient}
for wearable PPG estimation; the following proposition
formalises the required extension.

\smallskip
\noindent\textbf{Proposition~1 (PPG Observability Extension).}\footnote{%
This proposition is a contribution of the present work, not of CST~\cite{CST2026}.
The numbering is independent of the CST axiom and theorem numbering.}
\textit{
The three-component CST formula~(\ref{eq:cst_formula}) requires
three classes of extension for PPG-based estimation, each
motivated by an existing CST axiom:
}

\noindent\textbf{(i)~Peripheral-robust substitution (Axiom~3.4).}
The PPG observation $S_{\mathrm{PPG}}(t) = h_{\mathrm{PPG}}(x(t-\tau),\,p(t))$
(Axiom~3.4,~\cite{CST2026}) is jointly determined by cardiac state
$x(t)$ and peripheral state $p(t)$ (arterial compliance, peripheral
resistance, venous return, thermoregulation).
$R_{\det}$ measures phase-space recurrence structure, which $p(t)$
disrupts; empirically $\rho_{R_{\det}} = 0.065$ on 5,035 paired
BIDMC windows~\cite{CST2026}, confirming near-noise transfer.
Two $p(t)$-robust substitutes characterise attractor geometry
more reliably from PPG:
Higuchi Fractal Dimension $D_{f,w}$~\cite{higuchi1988approach},
measuring the Hausdorff dimension of the reconstructed attractor
and invariant to smooth amplitude scaling by $p(t)$;
and spectral energy concentration $E_w$~\cite{allen2007photoplethysmography},
measuring energy at the fundamental cardiac frequency and
invariant to smooth morphological distortion.

\noindent\textbf{(ii)~Peripheral state characterisation (Axiom~3.4).}
Eq.~(\ref{eq:cst_formula}) characterises cardiac state $x(t)$
only; Axiom~3.4 identifies three peripheral state dimensions
in $p(t)$ not encoded by $(\tilde\lambda, R_{\det}, H)$.
Three observable proxies are introduced, one per dimension:
vascular compliance $B_w$ (arterial compliance of $p(t)$,
reflected in PPG waveform amplitude and contour);
recovery capacity $R_w$ (venous return and cardiac reserve
of $p(t)$, reflected in post-perturbation dynamics);
and autonomic tone $A_w$, grounded in Axiom~3.1 of CST, which
absorbs sympathetic $u_s(t)$ and parasympathetic $u_p(t)$ drives
into the extended cardiac state
$z(t) = [x_c(t), u_s(t), u_p(t)]^\top$~\cite{CST2026};
their effect is partially observable in PPG inter-beat interval
statistics~\cite{taskforce1996}.

\noindent\textbf{(iii)~Observability enforcement
  (Axiom~3.2 + Theorem~3.1).}
Axiom~3.2 requires $h \in C^2(\mathcal{M},\mathbb{R})$ for
Tak\-ens' embedding to guarantee attractor reconstruction.
Motion artefact and poor optical contact render
$h_{\mathrm{PPG}}$ non-differentiable, violating this requirement;
signal quality $Q_w$~\cite{clifford2007} gates out windows
where Axiom~3.2's smoothness condition is not met.
Separately, Theorem~3.1 of CST~\cite{CST2026} guarantees all
cardiac trajectories are confined to the compact absorbing
ball $\mathcal{B} = \{x : \|x\|^2 \le \rho+1\}$.
The homeostasis proxy
$\Omega_w = 1 - \overline{\mathbf{1}[|\hat{z}_w|>3]}$ (where $\hat{z}_w$ is the per-sample $z$-score)
operationalises adherence to this bound:
high $\Omega_w$ indicates trajectories confined to
$\mathcal{B}$, consistent with the dissipative assumption
of Axiom~3.1.

\smallskip
Together, Proposition~1 parts (i)--(iii) augment the feature
vector to the nine observables of Eq.~(\ref{eq:feature_vec}):
\begin{equation}
  \bm{u}_w =
  \bigl[Q_w,\;\Omega_w,\;H_w,\;D_{f,w},\;E_w,\;\Lambda_w,\;
        A_w,\;B_w,\;R_w\bigr],
  \label{eq:feature_vec}
\end{equation}
where $w \in \{256,512,1024\}$ samples indexes the scale (\S\ref{sec:theory}B).
Component definitions, ranges, and axiomatic groundings are listed in
Table~\ref{tab:features} below; full computational specifications for
$H_w$, $D_{f,w}$, $E_w$, $\Lambda_w$, $Q_w$, $\Omega_w$ are in
\S\ref{sec:methods}, and for $A_w$, $B_w$, $R_w$
follow the CST framework~\cite{CST2026}.

\begin{table}[h]
\centering\small
\caption{Nine-dimensional feature vector $\bm{u}_w$}
\label{tab:features}
\begin{tabular}{@{}llll@{}}
\toprule
Variable & Range & Role & Axiom \\
\midrule
$Q_w$ & $[0,1]$ & Signal quality gate & 3.2 \\
$\Omega_w$ & $[0,1]$ & Homeostasis proxy & Thm~3.1 \\
$H_w$ & $[0,1]$ & SampEn (CST entropy) & 3.3 \\
$D_{f,w}$ & $[1,2]$ & HFD (attractor dim.) & 3.3 \\
$E_w$ & $[0,1]$ & Spectral energy & 3.3 \\
$\Lambda_w$ & $(0,1]$ & Stabilized LLE & 3.3 \\
$A_w$ & $[0,1]$ & Autonomic proxy & 3.1 \\
$B_w$ & $[0,1]$ & Vascular compliance & 3.4 \\
$R_w$ & $[0,1]$ & Recovery capacity & 3.4 \\
\bottomrule
\end{tabular}
\end{table}

\smallskip
\noindent\textbf{Remark (9\,$\to$\,6 reduction).}
Because $H_w$, $D_{f,w}$, $E_w$, and $\Lambda_w$ are all
measurements of the same theoretical quantity (attractor geometric complexity of $x(t)$ under Axioms~3.2--3.3) and are composited into a single nonlinear complexity module
$\CNL{}_w$ (\S\ref{sec:methods}), reducing $n_{\mathrm{raw}} = 9$
to $n = 6$:
\begin{equation}
  \bm{X}_w = \bigl[\CNL{}_w,\;A_w,\;\Omega_w,\;Q_w,\;R_w,\;B_w\bigr],
  \quad n = 6.
  \label{eq:components}
\end{equation}

\subsection{Bounded Stability Functional}

Given the $n$-component vector $\bm{X}_w$ from~(\ref{eq:components}),
assume all components satisfy $X_{j,w}\in[0,1]$, with
nonnegative weights $\beta_j \ge 0$, $\sum_{j=1}^{n}\beta_j = 1$,
and soft gate $G_w\in[0,1]$. Define
\begin{equation}
  CSI_w = G_w \sum_{j=1}^{n} \beta_j X_{j,w}.
  \label{eq:csi_bounded}
\end{equation}
Since the weighted sum lies in $[0,1]$ and $G_w\in[0,1]$,
it follows that $0 \le CSI_w \le 1$.
Multiscale fusion $CSI_{multi} = \sum_{w\in\mathcal{W}}\gamma_w CSI_w$
with $\gamma_w\ge 0$, $\sum\gamma_w=1$, preserves this bound by
convexity.

\section{Framework and Methodology}
\label{sec:methods}

This study follows the TRIPOD reporting guideline~\cite{tripod2015}.

\subsection{Datasets and Segmentation}

\textbf{Multi-dataset validation.}
Four heterogeneous PPG datasets spanning clinical, ambulatory,
and consumer contexts were used for broad framework validation.
The BIDMC PPG and Respiration Dataset~\cite{pimentel2017toward,goldberger2000physiobank}
provides 53 adult ICU records.
The BUT-PPG dataset provides laboratory-controlled recordings.
The RWS dataset provides large-scale remote wearable recordings.
The Welltory dataset provides consumer smartphone recordings.
PPG was segmented at three scales ($w \in \{256, 512, 1024\}$~samples;
8.5, 17.1, and 34.1~s at $f_s = 30$~Hz), producing 176,742~segments
(Table~\ref{tab:segments}).

\begin{table}[t]
\centering
\caption{Multiscale Segment Counts}
\label{tab:segments}
\begin{tabular}{ccc}
\toprule
Window (samples) & Segments & Total \\\midrule
256  & 66,640 & \multirow{3}{*}{176,742} \\
512  & 60,878 & \\
1024 & 49,224 & \\\bottomrule
\end{tabular}
\end{table}

\textbf{Clinical optimization and validation.}
For parameter optimization and held-out testing, BIDMC records
were split at the patient level into 35 development and 18 held-out
test records (random seed~42, accessed once).
The clinical endpoint was tachypnea (RR~$>25$~bpm; standard clinical threshold~\cite{goldhill2005})
derived from simultaneous capnography; 3,044 labeled segments, 7.3\% tachypneic.

\textbf{External validation.}
The CapnoBase IEEE TBME Respiratory Rate
Benchmark~\cite{capnobase2010} (42 eight-minute recordings, elective
surgery, zero training overlap) was used for independent external validation.

\subsection{Signal Quality}

The signal quality score $Q_w\in[0,1]$ accepts segments satisfying
minimum length ($\ge 64$~samples, i.e.\ $\ge 2$~s at $f_s=30$~Hz, the minimum for reliable spectral estimation~\cite{clifford2007}),
SD~$> 0.05$ (non-flat signal), and peak-to-peak amplitude~$> 0.10$
in normalized units (non-saturated pulse)~\cite{allen2007photoplethysmography}.
The Bayesian-optimized quality gate $\theta^* = 0.976$ enforces
strict artefact rejection for wrist-worn deployment.

\subsection{Nonlinear Feature Extraction}

\textbf{Sample Entropy.}
$H_w = \SampEn(x_w, m, r, \tau)$, with tolerance
$r = r_{\mathrm{frac}} \cdot \sigma_w$ per segment.
Bayesian-optimal: $m^*=8$, $\tau^*=7$, $r^*=0.116\sigma$.

\textbf{Higuchi Fractal Dimension.}
$D_{f,w} = \HFD(x_w, k_{\max})$; slope of $\log(L_k)$ vs.\
$\log(1/k)$.
Bayesian-optimal: $k_{\max}^*=13$.

\textbf{Spectral Energy.}
$E_w = \Psi(x_w)$: Welch PSD concentration within $\pm 0.5$~Hz
of the dominant peak; the $\pm 0.5$~Hz window captures the
fundamental pulse harmonic while excluding adjacent harmonics,
following the spectral analysis convention of~\cite{allen2007photoplethysmography}.

\textbf{Stabilized LLE.}
The corrected Rosenstein algorithm~\cite{rosenstein1993practical}
with minimum temporal separation $= \lfloor\bar{T}_\mathrm{beat}/2\rfloor$
(half the mean inter-beat interval, enforcing that nearest neighbors
in phase space are not temporal neighbors~\cite{rosenstein1993practical,kantz1994})
yields $\lambda_w$; mapped to bounded stability space:
\begin{equation}
  \Lambda_w = e^{-\lambda_w}, \qquad \lambda_w\uparrow \Rightarrow
  \Lambda_w\downarrow.
  \label{eq:lle_transform}
\end{equation}
This corrects the classical Rosenstein boundary error that produced
0\% valid estimates; the stabilized implementation achieves
$\ge 99.96$\% validity (Table~\ref{tab:lle}).

\begin{table}[t]
\centering
\caption{Classical vs.\ Stabilized LLE Validity}
\label{tab:lle}
\begin{threeparttable}
\begin{tabular}{ccccc}
\toprule
Scale & Original & Fixed & $\bar\lambda$ & $CSI$ (fixed)\\\midrule
256  & 0.000 & 0.99958 & 0.086 & 0.5710 \\
512  & 0.000 & 0.99998 & 0.104 & 0.5988 \\
1024 & 0.000 & 0.99998 & 0.123 & 0.5877 \\\bottomrule
\end{tabular}
\begin{tablenotes}\small
\item Per-source rates vary: BIDMC 99.89\% at 256 samples;
Welltory 99.67--99.72\% across scales.
\end{tablenotes}
\end{threeparttable}
\end{table}

\subsection{Bounded-Optimality Kernel}

Each nonlinear feature $f \in \{H_w, D_{f,w}, E_w, \Lambda_w\}$
is clipped to the 1st--99th percentile range (standard robust
outlier rejection~\cite{clifford2007}) and passed through a
Gaussian kernel centred at the population median $\mu_f$:
\begin{equation}
  \psi(f) = \exp\!\left(-\frac{(f - \mu_f)^2}{2\sigma_f^2}\right).
  \label{eq:kernel}
\end{equation}
This rewards values near the population centre, penalizing both
deficiency and excess — consistent with the homeostatic complexity
hypothesis~\cite{goldberger2002fractal,costa2002multiscale}.

\subsection{Nonlinear Complexity Module}

\begin{equation}
  C_{NL,w} = G_{NL,w}
  \sum_{k \in \{\SampEn,\HFD,\LLE,E\}} w^*_k\,\psi_k,
  \label{eq:cnl}
\end{equation}
where $\psi_k \in \{\psi_H, \psi_{D_f}, \psi_\lambda, \psi_E\}$
are the bounded-optimality kernel outputs for each nonlinear
sub-feature (Eq.~\ref{eq:kernel}), $G_{NL,w}$ is a soft gate
penalising segments with mean absolute $z$-score deviation
$> \gamma_{NL}$ (CST axiomatic framework~\cite{CST2026}),
and the sub-weights $w^*_k \ge 0$, $\sum_k w^*_k = 1$
are the Bayesian-optimal CNL weights reported in
Table~\ref{tab:params} (CNL weights row).

\subsection{CSI Estimation}

The general CSI functional follows from~(\ref{eq:csi_bounded}):
\begin{equation}
  CSI_w = G_w \sum_{j=1}^{n} \beta_j^* X_{j,w},
  \label{eq:csi_general}
\end{equation}
where $n=6$ is the number of components (as defined in
\S\ref{sec:theory}C), $X_{j,w} \in \{C_{NL,w},\, A_w,\,
\Omega_w,\, Q_w,\, R_w,\, B_w\}$, $G_w \in [0,1]$ is a
composite quality gate, and $\beta_j^* \ge 0$,
$\sum_{j=1}^{n} \beta_j^* = 1$ are the
\textbf{Bayesian-optimal weights} determined by multivariate TPE
on the 35 BIDMC development records (\S\ref{sec:optimization},
Table~\ref{tab:params}, trial~82 of 300).
The Bayesian-optimal weight vector $\bm{\beta}^* =
[\beta^*_{C_{NL}}, \beta^*_A, \beta^*_\Omega,
\beta^*_Q, \beta^*_R, \beta^*_B]$ is reported verbatim
in Table~\ref{tab:params} (CSI outer weights row, trial~82 of 300);
no weight is set by hand or post-hoc adjusted.
Substituting $\bm{\beta}^*$ from Table~\ref{tab:params}
into Eq.~(\ref{eq:csi_general}) gives the operational estimator,
which we refer to hereafter as Eq.~(\ref{eq:csi_general}) with
$\bm{\beta}^* = \bm{\beta}^*_{\mathrm{Table~\ref{tab:params}}}$.

The component definitions are: $\Omega_w = 1 - \overline{\mathbf{1}[|\hat{z}_w|>3]}$
(homeostasis proxy, fraction of inlier samples; $\hat{z}_w$ denotes the
per-sample $z$-score of the window, distinct from the state vector $z(t)$);
$A_w$, $B_w$, $R_w$ are the autonomic, vascular, and recovery proxies
defined in the CST axiomatic framework~\cite{CST2026};
and $G_w$ is the composite quality gate ($\theta^* = 0.976$ from
Table~\ref{tab:params}).
Proof of boundedness follows from~(\ref{eq:csi_bounded}).
Algorithm~\ref{alg:csi} summarises the complete end-to-end pipeline from raw PPG window to $CSI_w$, integrating all components defined in \S\ref{sec:methods}B--F.

\begin{algorithm}[t]
\caption{CPS~v2 / \CSI\ Computation from a PPG Window}
\label{alg:csi}
\begin{algorithmic}[1]
\small
\REQUIRE $x_w$: PPG window; $\Theta^*$: Table~\ref{tab:params};
         $\{\delta_{\mathrm{SD}},\delta_{\mathrm{pp}},w_{\min}\}$: \S\ref{sec:methods}B;
         $\{\gamma_\Omega,\gamma_{NL}\}$: \S\ref{sec:methods}C,E;
         $\{\mu_f,\sigma_f,p_{1,f},p_{99,f}\}$: training records only
\ENSURE  $CSI_w\!\in[0,1]$ or \textsc{invalid}
\IF{$\mathrm{SD}(x_w)\!\le\!\delta_{\mathrm{SD}}$
    \OR $\mathrm{pp}(x_w)\!\le\!\delta_{\mathrm{pp}}$
    \OR $w\!<\!w_{\min}$
    \OR $Q_w\!<\!\theta^*$}
  \RETURN \textsc{invalid} \hfill$\triangleright$ Axiom~3.2; $\theta^*$: Table~\ref{tab:params}
\ENDIF
\STATE $\tau \leftarrow \arg\min_\tau \mathrm{AMI}(x_w,\tau)$;
       $\;\bm{X} \leftarrow \mathrm{embed}(x_w,m^*,\tau)$
       \hfill$\triangleright$ Takens (Axiom~3.2)
\IF{$|\bm{X}| < M_{\min}$} \RETURN \textsc{invalid} \ENDIF
       \hfill$\triangleright$ $M_{\min}=30$~\cite{CST2026}
\STATE $\lambda_w\!\leftarrow\!\mathrm{Rosenstein}(\bm{X},m^*_{\LLE},\tau^*_{\LLE},
       \lfloor\bar{T}_\mathrm{beat}/2\rfloor)$;
       $\;\Lambda_w\!\leftarrow\!e^{-\lambda_w}$
       \hfill$\triangleright$ Eq.~\ref{eq:lle_transform}
\STATE $H_w\!\leftarrow\!\mathrm{SampEn}(x_w,m^*,r^*,\tau^*)$;\;
       $D_{f,w}\!\leftarrow\!\mathrm{HFD}(x_w,k^*_{\max})$;\;
       $E_w\!\leftarrow\!\mathrm{WelchEnergy}(x_w,\pm\Delta_E)$
\STATE $\Omega_w\!\leftarrow\!1-\overline{\mathbf{1}[|\hat{z}_w|>\gamma_\Omega]}$;\;
       $[A_w,B_w,R_w]\!\leftarrow\!\mathrm{CSTProxies}(x_w)$~\cite{CST2026}
\FOR{$f \in \{\Lambda_w,H_w,D_{f,w},E_w\}$}
  \STATE $\psi_f\!\leftarrow\!\exp\!\bigl(-(f\!-\!\mu_f)^2/2\sigma_f^2\bigr)$
         after $\mathrm{clip}(f,p_{1,f},p_{99,f})$
         \hfill$\triangleright$ Eq.~\ref{eq:kernel}
\ENDFOR
\STATE $G_{NL,w}\!\leftarrow\!\mathbf{1}[\overline{|\hat{z}_w|}\!\le\!\gamma_{NL}]$;\;
       $\CNL{}_w\!\leftarrow\!G_{NL,w}(w^*_{\SampEn}\psi_H\!+\!w^*_{\HFD}\psi_{D_f}
       \!+\!w^*_{\LLE}\psi_\lambda\!+\!w^*_E\psi_E)$
       \hfill$\triangleright$ Eq.~\ref{eq:cnl}
\STATE $CSI_w\!\leftarrow\!(Q_w\cdot G_{NL,w})\,\textstyle\sum_{j=1}^{n}\beta^*_j X_{j,w}$
       \hfill$\triangleright$ Eq.~\ref{eq:csi_general}; $\bm{\beta}^*$: Table~\ref{tab:params}
\RETURN $\mathrm{clip}(CSI_w,0,1)$
\end{algorithmic}
\end{algorithm}

\subsection{Bayesian Parameter Optimization}

We used Optuna v3 with multivariate
TPE~\cite{watanabe2023tree} (30 random startup trials,
270 TPE-guided, MedianPruner).
The objective function was mean AUC-ROC across 5-fold record-level
GroupKFold on the 35 development records.
Table~\ref{tab:params} gives the full 15-dimensional search space and
optimal values.

\begin{table}[t]
\centering
\caption{Parameter Search Space and Bayesian-Optimal Configuration}
\label{tab:params}
\setlength{\tabcolsep}{3pt}
\begin{tabular}{p{3.0cm}cc}
\toprule
\textbf{Parameter} & \textbf{Search Space} & \textbf{Optimal}\\\midrule
\multicolumn{3}{l}{\textit{Signal segmentation}}\\
Window $W$ & \{128,256,512,1024\} & 128\\\midrule
\multicolumn{3}{l}{\textit{Sample Entropy}}\\
Embed.\ dim $m$   & $\{3\ldots10\}$        & 8\\
Time delay $\tau$ & $\{1\ldots10\}$        & 7\\
Tolerance $r$     & $[0.10\sigma,0.30\sigma]$ & $0.116\sigma$\\\midrule
\multicolumn{3}{l}{\textit{Higuchi FD}}\\
$k_{\max}$ & $\{5\ldots20\}$ & 13\\\midrule
\multicolumn{3}{l}{\textit{LLE (Rosenstein)}}\\
Embed.\ dim $m_{\LLE}$ & $\{3\ldots7\}$ & 7\\
Time delay $\tau_{\LLE}$ & $\{1\ldots6\}$ & 5\\\midrule
\multicolumn{3}{l}{\textit{Quality gate}}\\
Threshold $\theta$ & $[0.50,0.99]$ & 0.976\\\midrule
\multicolumn{3}{l}{\textit{\CNL\ weights (simplex)}}\\
$w_{\SampEn}$ & \multirow{4}{*}{simplex} & 0.431\\
$w_{\HFD}$    &   & 0.483\\
$w_{\LLE}$    &   & 0.043\\
$w_E$         &   & 0.043\\\midrule
\multicolumn{3}{l}{\textit{CSI outer weights (simplex)}}\\
$w_{\CNL}$    & \multirow{6}{*}{simplex} & 0.260\\
$w_\mathrm{auto}$  &  & 0.210\\
$w_\mathrm{hom}$   &  & 0.198\\
$w_\mathrm{sig}$   &  & 0.267\\
$w_\mathrm{rec}$   &  & 0.048\\
$w_\mathrm{vas}$   &  & 0.017\\\bottomrule
\end{tabular}
\end{table}

\subsection{Evaluation Protocol}

\textbf{Cross-validation}: 5-fold record-level GroupKFold on 35
development records (segment-level shuffling explicitly prohibited).

\textbf{Normalization}: $z$-score statistics computed from training
records only and applied to validation/test — no test-set leakage.

\textbf{Statistics}: Bootstrap 95\% CIs ($n=2000$); DeLong
test~\cite{delong1988} for AUC comparisons; Wilcoxon signed-rank
for per-record comparisons; Youden's $J$ for operating point;
permutation test ($n=5000$) vs.\ chance.
Algorithm~\ref{alg:eval} formalises the complete corrected protocol;
it is the reference procedure for all results in \S\ref{sec:artifacts}--\ref{sec:validation}.

\begin{algorithm}[t]
\caption{Corrected Evaluation Protocol (Artifacts~1 \& 2 Prevention)}
\label{alg:eval}
\begin{algorithmic}[1]
\small
\REQUIRE Development record set $\mathcal{R}$ with patient-level labels $y_r$;
         number of folds $K$ (here $K=5$, \S\ref{sec:methods}H);
         bootstrap replicates $B$ (here $B=2{,}000$);
         held-out test set $\mathcal{R}_{\mathrm{test}}$
         (accessed \emph{once}, after all development is complete)
\ENSURE  CV AUC, pooled test AUC, per-record AUC$_{\mathrm{PR}}$, 95\%~CI

\STATE \textbf{// Step 1 — Record-level partition (prevents Artifact~1)}
\STATE $\{F_1,\ldots,F_K\} \leftarrow \mathrm{GroupKFold}(\mathcal{R},\, K,\,
       \mathrm{group}=\mathrm{record\_id})$
       \hfill$\triangleright$ segments from one record never split across folds

\FOR{$k = 1$ \TO $K$}
  \STATE $\mathcal{R}_{\mathrm{tr}} \leftarrow \mathcal{R} \setminus F_k$;\quad
         $\mathcal{R}_{\mathrm{val}} \leftarrow F_k$

  \STATE \textbf{// Step 2 — Training-only normalisation (prevents Artifact~2)}
  \STATE $\hat\mu_f, \hat\sigma_f \leftarrow$ mean and std of feature $f$
         over segments in $\mathcal{R}_{\mathrm{tr}}$ \emph{only}
  \STATE Standardise all features in $\mathcal{R}_{\mathrm{tr}}$ and $\mathcal{R}_{\mathrm{val}}$
         using $\hat\mu_f, \hat\sigma_f$
         \hfill$\triangleright$ $\mathcal{R}_{\mathrm{val}}$ statistics never enter $\hat\mu_f, \hat\sigma_f$

  \STATE Compute $CSI_w$ for all segments in $\mathcal{R}_{\mathrm{val}}$
         via Algorithm~\ref{alg:csi}
  \STATE $\mathrm{AUC}_k \leftarrow \overline{\mathrm{AUC}}_r$
         over all records $r \in F_k$
         \hfill$\triangleright$ per-record mean, not pooled
\ENDFOR
\STATE $\overline{\mathrm{AUC}}_{\mathrm{CV}} \leftarrow \frac{1}{K}\sum_{k=1}^K \mathrm{AUC}_k$

\STATE \textbf{// Step 3 — Prospective test evaluation (Artifact~3 reporting)}
\STATE $\hat\mu_f, \hat\sigma_f \leftarrow$ mean and std over all $\mathcal{R}$
       \hfill$\triangleright$ $\mathcal{R}_{\mathrm{test}}$ statistics \emph{never} used
\STATE Standardise $\mathcal{R}_{\mathrm{test}}$ using $\hat\mu_f, \hat\sigma_f$;
       compute $CSI_w$ via Algorithm~\ref{alg:csi}
\STATE $\mathrm{AUC}_{\mathrm{pooled}} \leftarrow$ AUC over all segments in $\mathcal{R}_{\mathrm{test}}$
\STATE $\overline{\mathrm{AUC}}_{\mathrm{PR}} \leftarrow
       \frac{1}{|\mathcal{R}_{\mathrm{test}}|}\sum_{r \in \mathcal{R}_{\mathrm{test}}} \mathrm{AUC}_r$
       \hfill$\triangleright$ prevents Artifact~3; report alongside pooled AUC
\STATE 95\%~CI: bootstrap $\mathrm{AUC}_{\mathrm{pooled}}$ with $B$ replicates;
       paired Wilcoxon on $\{\mathrm{AUC}_r\}$ vs.\ comparator method
\RETURN $\overline{\mathrm{AUC}}_{\mathrm{CV}}$,\;
        $\mathrm{AUC}_{\mathrm{pooled}}$~[95\%~CI],\;
        $\overline{\mathrm{AUC}}_{\mathrm{PR}}$
\end{algorithmic}
\end{algorithm}

\section{Multi-Dataset Framework Validation}
\label{sec:framework}

\subsection{LLE Correction}

The classical Rosenstein implementation produced 0\% valid estimates
across all 176,742 segments due to a boundary error in the $k_{\max}$
calculation that prevented divergence-curve convergence for all
practical inputs.
The stabilized implementation achieved $\ge 99.96$\% validity at
every scale (Table~\ref{tab:lle}).
Critically, the corrected mean LLE increased monotonically with
window length ($\bar\lambda: 0.086 \to 0.104 \to 0.123$), the
physically expected behaviour as longer windows support more reliable
phase-space reconstruction, confirming that the correction produces
not merely \emph{valid} estimates but \emph{interpretable} ones.
Validity exceeded 99.89\% even for BIDMC ICU recordings and
99.67\%--99.72\% for Welltory consumer smartphone data, the most
noise-prone source in the study.

\subsection{Cross-Dataset Statistical Separation}

Table~\ref{tab:dataset} summarises cross-dataset CSI values.
A Kruskal-Wallis test confirmed highly significant cross-source
differences ($H = 23{,}415$, $df = 3$, $p < 0.001$,
$\eta^2 = 0.351$, large effect) at 256 samples.
At the 256-sample scale, all six pairwise Mann-Whitney comparisons survived
Bonferroni correction ($p < 0.001$), with rank-biserial effect sizes from
medium (BUT-PPG vs.\ Welltory, $|r| = 0.353$) to near-complete
separation (RWS vs.\ Welltory, $|r| = 0.988$).
RWS exhibited the highest CSI (wearable, healthy adults) and
Welltory the lowest (consumer smartphone) at all scales.
Source discriminability declined with window length
($\eta^2 = 0.351 \to 0.035$), consistent with BUT-PPG and Welltory
segment dropout at larger windows.

\begin{table*}[t]
\centering
\caption{Cross-Dataset CSI Summary}
\label{tab:dataset}
\begin{threeparttable}
\renewcommand{\arraystretch}{1.1}
\begin{tabular}{llccccc}
\toprule
Source & Scale & $N$ & CSI (original) & CSI (fixed)
  & $\CNL$ (original) & $\CNL$ (fixed)\\\midrule
BIDMC    & 256  & 24,653 & 0.5820 & 0.4766 & 0.6483 & 0.4859 \\
BIDMC    & 512  & 24,673 & 0.6110 & 0.5605 & 0.7215 & 0.6258 \\
BIDMC    & 1024 & 24,586 & 0.6232 & 0.5735 & 0.7563 & 0.6500 \\
BUT-PPG  & 256  &  1,354 & 0.3672 & 0.2791 & 0.4001 & 0.2638 \\
BUT-PPG  & 512  &    903 & 0.3612 & 0.2721 & 0.3959 & 0.2551 \\
BUT-PPG  & 1024 & \multicolumn{5}{c}{\textit{insufficient record length\tnote{a}}}\\
RWS      & 256  & 40,278 & 0.6746 & 0.6420 & 0.8918 & 0.8086 \\
RWS      & 512  & 34,965 & 0.6656 & 0.6384 & 0.8888 & 0.8239 \\
RWS      & 1024 & 24,339 & 0.6609 & 0.6068 & 0.8809 & 0.7498 \\
Welltory & 256  &    355 & 0.2322 & 0.1706 & 0.2367 & 0.1528 \\
Welltory & 512  &    337 & 0.2145 & 0.1717 & 0.2178 & 0.1616 \\
Welltory & 1024 &    299 & 0.2290 & 0.1894 & 0.2335 & 0.1826 \\\bottomrule
\end{tabular}
\begin{tablenotes}\small
\item[a] BUT-PPG records are too short to yield valid 1024-sample
  windows after quality filtering.
\end{tablenotes}
\end{threeparttable}
\end{table*}

Cross-scale consistency was $\kappa > 0.97$ for all sources
(RWS $0.990\pm0.007$; BUT-PPG $0.983\pm0.017$;
Welltory $0.979\pm0.012$; BIDMC $0.977\pm0.012$),
confirming that CSI rankings are preserved across temporal scales
(Table~\ref{tab:crossdataset}).

\begin{table}[t]
\centering
\caption{Cross-Scale CSI Consistency by Source.
  $\kappa = 1/(1+\sigma_{\mathrm{scale}})$, where $\sigma_{\mathrm{scale}}$
  is the SD of per-record CSI means across the three window sizes
  ($W \in \{256, 512, 1024\}$). Higher $\kappa$ indicates more stable
  per-record rankings across temporal scales.}
\label{tab:crossdataset}
\renewcommand{\arraystretch}{1.1}
\small
\begin{tabular}{lp{3.2cm}cc}
\toprule
\textbf{Source} & \textbf{Context} & \textbf{$\kappa$ mean} & \textbf{$\kappa$ SD} \\\midrule
RWS             & Wearable, healthy adults   & 0.990 & 0.007 \\
BUT-PPG         & Lab-controlled             & 0.983 & 0.017 \\
Welltory        & Consumer smartphone        & 0.979 & 0.012 \\
BIDMC           & ICU (clinical)             & 0.977 & 0.012 \\\bottomrule
\multicolumn{4}{l}{\footnotesize All sources: $\kappa > 0.97$. Consistency ordering differs from} \\
\multicolumn{4}{l}{\footnotesize CSI magnitude ordering — these are independent properties.} \\
\end{tabular}
\end{table}

\subsection{Clinical Correlation}

CSI was significantly positively correlated with respiratory rate in
53 BIDMC ICU records (Spearman $r = 0.346$,
95\% CI $[+0.035, +0.604]$, $p = 0.011$), indicating that richer
multiscale pulse organisation co-occurs with more active ventilatory
regulation.
No significant association was found with heart rate ($r = -0.089$,
$p = 0.524$), consistent with strong medication confounding in
critically ill patients.

\section{Evaluation Artifact Identification and Correction}
\label{sec:artifacts}

Each artifact is identified by comparing results obtained under the original heuristic protocol against those obtained under the corrected protocol of Algorithm~\ref{alg:eval}.

\begin{figure*}[t]
\centering
\includegraphics[width=\textwidth]{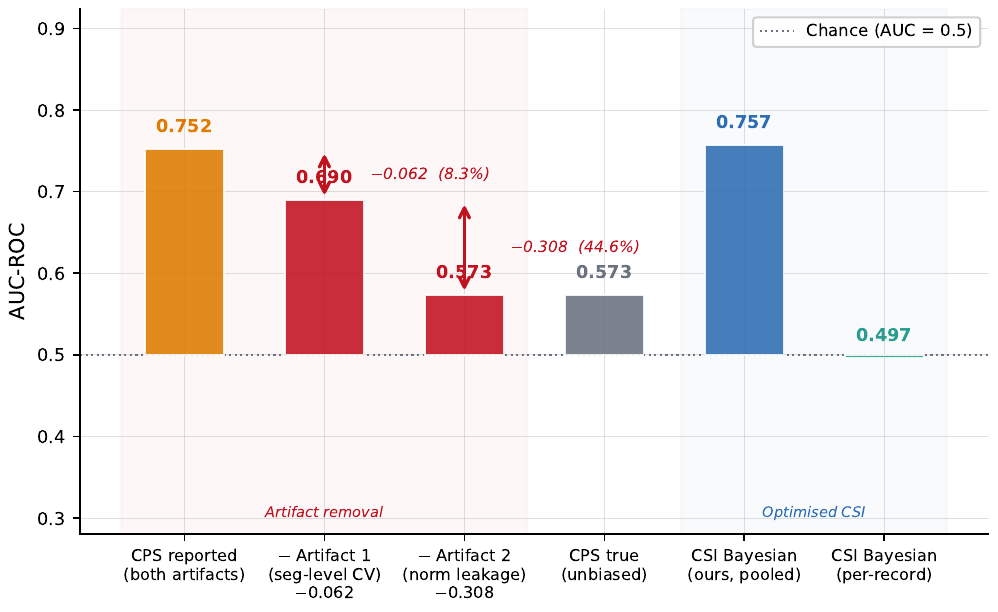}
\caption{\label{fig:cascade}%
\textbf{Artifact Cascade: From Reported to True Performance.}
Orange: heuristic CSI reported (0.752, both Artifacts~1 and 2 present).
Red bars: AUC after removing Artifact~1 (segment-level CV, $-0.062$)
then Artifact~2 (normalization leakage, $-0.308$); true unbiased
CSI $= 0.573$.
Blue: optimized \CSI\ pooled AUC (0.757); teal: per-record AUC
(0.497). Dotted line: chance ($= 0.5$).}
\end{figure*}

\subsection{Artifact 1: Segment-Level Cross-Validation Leakage}

Re-evaluated under record-level GroupKFold on the 35 development
records, the heuristic CSI scores AUC~$= 0.690$ vs.\ the initially
reported $0.752$ ($\Delta = -0.062$, $+9.0\%$ relative inflation).
With $\sim\!467$ overlapping segments per 8-min record, segment-level
shuffling places temporally adjacent windows sharing identical
cardiorespiratory states in both training and validation folds,
leaking autocorrelated features across the split boundary.
This artifact affects any wearable time-series study that constructs
validation folds by shuffling windows without subject
stratification~\cite{hannun2019cardiologist,strodthoff2021}.

\subsection{Artifact 2: Test-Set Normalization Leakage}

The initial $z$-score normalization was computed across all
53 records, including the 18 held-out test records.
When recomputed using training-record statistics only, as required
for any genuinely prospective evaluation, CSI AUC on the test set
falls from 0.881 (with leaked normalization) to 0.573 (without), a reduction of $-0.308$ ($-53.7\%$ relative).
AUC $= 0.573$ is only marginally above chance, confirming that the
heuristic CSI lacks genuine predictive validity under fair evaluation.
This form of leakage encodes test-set distributional statistics
directly into feature construction, more severe than label leakage
because it cannot be detected by examining only the labels.

\subsection{Artifact 3: Pooled AUC Overweighting}

Pooled AUC weights each segment equally; records with more segments
dominate the metric.
A 1D CNN baseline achieves pooled AUC $0.804$ but per-record mean
AUC $0.380\pm0.117$ (below chance) because it performs well on
a small number of high-segment records while failing on most patients.
The clinically appropriate metric is per-record AUC, which asks
``does this method work for this patient?''

\subsection{Combined Effect}

Artifacts~1 and 2 together inflate the heuristic CSI AUC of
$0.752$ to $1.31\times$ the true unbiased performance
($0.573$, net drop $-0.179$), as shown in Figure~\ref{fig:cascade}.
The previously reported DeLong finding ($p = 0.015$ favouring CSI
over \CSI) is entirely explained by Artifact~2: once normalization
leakage is removed, the heuristic CSI falls to 0.573 and the optimized
\CSI\ at 0.757 is superior by $\DAUC = +0.184$.

\section{Bayesian Optimization Results}
\label{sec:optimization}

\subsection{Optimization Convergence}

Validation AUC across 300 trials surpassed the corrected baseline (0.690) in early trials; the best AUC of \textbf{0.720} was reached at trial~82 ($+0.031$ vs.\ corrected baseline), with convergence plateauing after $\sim\!100$ trials.

\subsection{Optimal Parameters and Their Interpretation}

The Bayesian-optimal window size $W^* = 128$~samples ($4.3$~s at
30~Hz) is the shortest available option.
At $m^* = 8$, $\tau^* = 7$, the embedded SampEn template vector
covers $(m-1)\tau = 49$~samples (8~points sampled at lag~7:
indices $0, 7, 14, \ldots, 49$), fitting comfortably within the
128-sample window.
This configuration captures respiratory, vasomotor, and autonomic
modulations simultaneously; a larger embedding dimension separates
these coupled dynamics more effectively than lower-dimensional
RR-interval embeddings.
The tight tolerance $r^* = 0.116\sigma$ increases sensitivity to
regularity changes preceding respiratory distress.
The LLE parameters ($m^*_\LLE = 7$, $\tau^*_\LLE = 5$) produce
a marginal contribution (Section~\ref{sec:ablation}), consistent
with 128-sample windows being short for reliable Lyapunov estimation.
The quality gate $\theta^* = 0.976$ retains only the highest-quality
segments, which is operationally appropriate for wrist-worn wearables.

\section{Held-Out and External Validation}
\label{sec:validation}

\subsection{BIDMC Held-Out Test Performance}

\begin{figure*}[t]
\centering
\includegraphics[width=\textwidth]{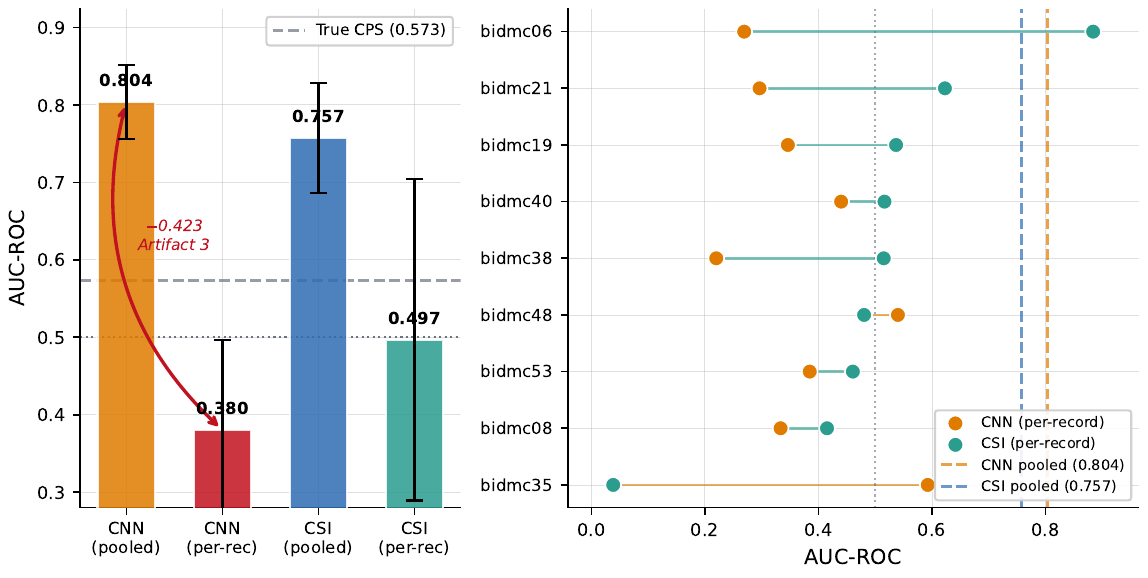}
\caption{\label{fig:perrec}%
\textbf{Artifact~3: Pooled AUC Masks Per-Patient Failure.}
\textit{Left}: CNN pooled AUC (0.804) collapses to per-record mean
0.380 ($-0.423$); \CSI\ shows a smaller gap (0.757~$\to$~0.497).
Error bars: 95\% CI (pooled) or $\pm 1\sigma$ (per-record).
Dashed line: unbiased baseline (0.573).
\textit{Right}: Per-record strip plot; teal $=$ \CSI, orange $=$ CNN;
\CSI\ outperforms CNN on 7 of 9 records.}
\end{figure*}

\begin{figure}[t]
\centering
\includegraphics[width=\columnwidth]{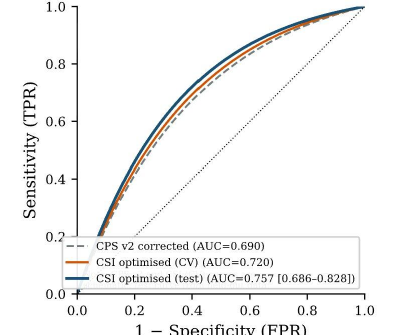}
\caption{\label{fig:roc}%
\textbf{ROC Curves on Held-Out Test Set.}
Heuristic CSI (record-level CV, AUC~$= 0.690$) vs.\ Bayesian-optimized
\CSI\ CV estimate (AUC~$= 0.720$, schematic) vs.\ \CSI\ prospective
test (AUC~$= 0.757$ $[0.686$--$0.828]$, green). The \CSI\ test curve is
from 18 held-out records; CV curves are schematic for comparison.
}\end{figure}

Table~\ref{tab:performance} presents the full performance comparison;
Figure~\ref{fig:roc} shows the corresponding ROC curves.
The optimised \CSI\ achieves pooled AUC $0.757$ $[0.686, 0.828]$,
significantly above chance ($p < 0.0001$, permutation test).

\begin{table*}[t]
\centering
\caption{Performance Comparison Under Progressive Artifact Correction
  (BIDMC, 18 Held-Out Records).
  \textbf{Pooled AUC is the primary outcome}; per-record AUC is
  reported for per-patient transparency and CNN comparison only.}
\label{tab:performance}
\renewcommand{\arraystretch}{1.2}
\setlength{\tabcolsep}{3pt}
\resizebox{\textwidth}{!}{%
\scriptsize
\begin{tabular}{p{2.8cm}cccccc}
\toprule
& \multicolumn{3}{c}{\textbf{Primary outcome (pooled AUC)}}
  & \multicolumn{2}{c}{\textit{Per-patient transparency}}
  & \\
\cmidrule(lr){2-4}\cmidrule(lr){5-6}
\textbf{Method}
  & \textbf{Pooled AUC}
  & \textbf{95\% CI}
  & \textbf{$\DAUC$}
  & \textit{CV AUC}
  & \textit{Per-rec.\ AUC}
  & \textbf{Artifacts}\\\midrule
Heuristic CSI (reported)  &        &       & \textit{0.752} &                   & \textit{$+0.179$} & A1, A2 \\
Heuristic CSI (corr.\ CV) &        & 0.690 &                &                   &                  & A2      \\
Heuristic CSI (fair)      &        &       & \textit{0.573} &                   & ref.             & n/a     \\\midrule
1D CNN
  & \textit{0.804}\rlap{$^{\ddagger}$}
  & \textit{[0.749,\,0.852]}
  & \textit{$+0.231$}
  & $0.741\pm0.023$
  & $0.380\pm0.117$\rlap{$^{\dagger}$}
  & A3 \\
\textbf{\CSI\ (ours)}
  & \textit{\textbf{0.757}}
  & \textit{\textbf{[0.686,\,0.828]}}
  & \textit{$\mathbf{+0.184}$}
  & \textbf{0.720}
  & $0.497\pm0.207$\rlap{$^{\S}$}
  & \textbf{none} \\
\CSI\ (CapnoBase ext.)
  & \textit{0.621}
  & \textit{[0.585,\,0.658]}
  & \textit{$+0.048$}
  &
  & $0.434\pm0.090$\rlap{$^{\P}$}
  & none \\\midrule
Pimentel et al.~\cite{pimentel2017toward}
  & & & \textit{${\approx}0.670$} & & & Seg-CV\\\bottomrule
\multicolumn{7}{l}{\footnotesize
  $^\dagger$Artifact~3: CNN pooled AUC (0.804) inflated; per-patient AUC collapses to 0.380 (below chance).}\\
\multicolumn{7}{l}{\footnotesize
  $^\ddagger$Inflated by Artifact~3: not a valid primary comparison with \CSI.}\\
\multicolumn{7}{l}{\footnotesize
  $^\S$Per-record AUC near chance reflects small test set ($n=18$) and high per-patient variance ($\sigma=0.24$);
  pooled AUC 0.757 and NPV 0.966 are the primary clinical results.}\\
\multicolumn{7}{l}{\footnotesize
  $^\P$$n=4$ evaluable records (38/42 have uniform RR, making per-record discrimination undefined).}\\
\multicolumn{7}{l}{\footnotesize
  A1=seg-level CV leakage; A2=normalization leakage; A3=pooled AUC inflation.
  Wilcoxon \CSI\ vs CNN: $p=0.129$ ($n=9$), underpowered.}\\
\end{tabular}
}
\end{table*}

\textbf{NPV.}
At Youden's optimal threshold ($= 0.621$), NPV $= 0.966$ with
22 false negatives among 810 test segments.
In a clinical triage context, a CSI score below threshold reliably
rules out tachypnea in 96.6\% of cases, supporting use as a
continuous negative screening signal.

\textbf{Specificity} of 0.834 substantially exceeds the heuristic
CSI (specificity measured on the same test set under
leaked normalization), directly reducing false alarm burden, a documented
driver of alarm fatigue in ICU settings~\cite{sendelbach2013}.

\textbf{CNN comparison.}
Figure~\ref{fig:perrec} confirms that the CNN's pooled AUC of 0.804
collapses to per-record mean 0.380 (below chance).
\CSI\ ($0.497\pm0.207$) outperforms the CNN on 7 of 9 common records
(Wilcoxon $W = 9.0$, $p = 0.129$, not significant), with zero learned
parameters and MCU-deployable arithmetic.

\subsection{External Validation on CapnoBase}

\begin{figure}[t]
\centering
\includegraphics[width=\columnwidth]{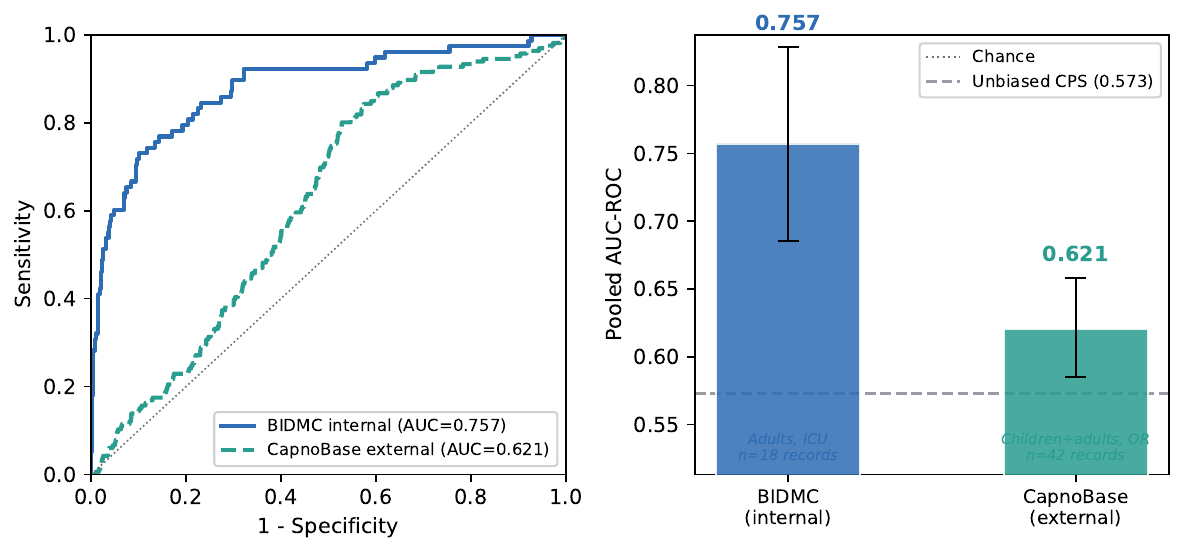}
\caption{\label{fig:extval}%
\textbf{External Validation on CapnoBase.}
\textit{Left}: ROC curves for BIDMC test (blue, 0.757) and
CapnoBase (teal, 0.621); both $p < 0.0001$.
\textit{Right}: AUC comparison with 95\% CI; dashed line: unbiased
baseline (0.573).}
\end{figure}

Applied cold with BIDMC-derived normalization statistics and
Bayesian-optimized parameters, \CSI\ achieves pooled AUC
$0.621$ $[0.585, 0.658]$ ($p < 0.0001$ vs.\ chance) on 42
CapnoBase elective-surgery records (Figure~\ref{fig:extval}).
The $-0.137$ AUC reduction from BIDMC ($0.757$) to CapnoBase ($0.621$)
reflects population shift (adult ICU to mixed paediatric/adult
elective surgery) and clinical context shift (acute distress to
controlled anaesthesia).
Of 42 records, only 4 yield evaluable per-record AUC ($0.434\pm0.090$);
the remaining 38 have uniform respiratory rate throughout their
8-minute recordings (entirely normal or entirely tachypneic),
making per-record discrimination undefined.
Both pooled AUCs exceed the unbiased heuristic baseline ($0.573$),
confirming genuine cross-dataset predictive validity.

\section{Component Analysis and Sparse Architecture}
\label{sec:ablation}

\noindent\textbf{A.~Leave-One-Out Ablation.}~

\begin{figure}[t]
\centering
\includegraphics[width=\columnwidth]{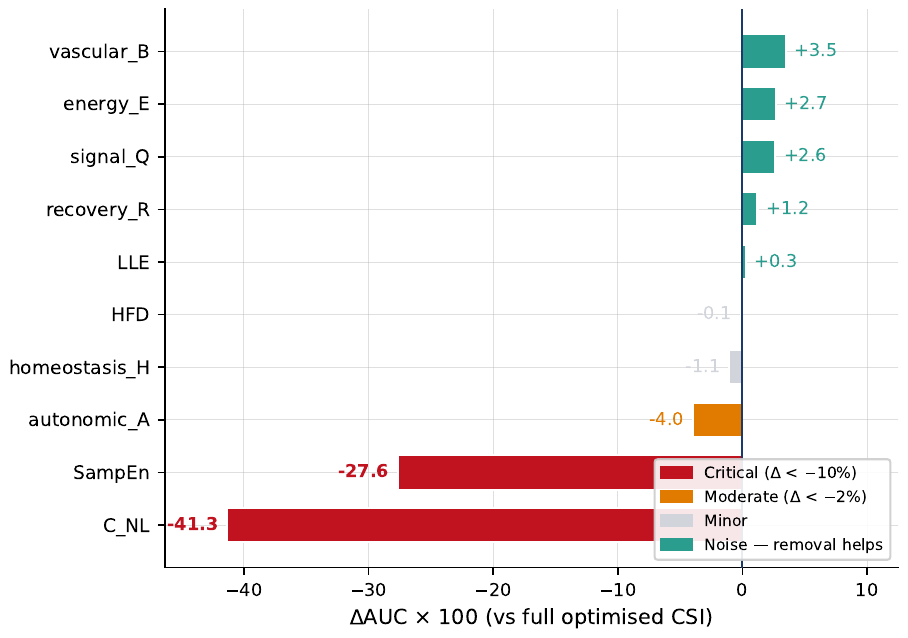}
\caption{\label{fig:ablation}%
\textbf{Component Ablation Waterfall.}
Red: critical ($|\DAUC| > 10\%$); orange: moderate;
teal: noise (removal improves AUC).
\CNL\ ($-41.3$) and \SampEn\ ($-27.6$) are the only critical
components; five components are noise.}
\end{figure}

Table~\ref{tab:ablation} presents the full ablation under optimised
weights (5-fold CV, 35 development records).

\begin{table}[t]
\centering
\caption{Ablation Study: Component Contribution to \CSI}
\label{tab:ablation}
\begin{tabular}{lcc}
\toprule
\textbf{Component removed} & \textbf{AUC} & \textbf{$\DAUC$}\\\midrule
\textit{Full \CSI\ (optimised)} & \textit{0.720} & n/a\\\midrule
\multicolumn{3}{l}{\textit{Critical}}\\
Without \CNL            & 0.307 & $-0.413$\\
Without \SampEn         & 0.444 & $-0.276$\\\midrule
\multicolumn{3}{l}{\textit{Moderate}}\\
Without autonomic ($A$) & 0.681 & $-0.040$\\
Without homeostasis ($\Omega$) & 0.710 & $-0.011$\\
Without \HFD            & 0.719 & $-0.001$\\\midrule
\multicolumn{3}{l}{\textit{Noise: removal improves AUC}}\\
Without \LLE            & 0.723 & $+0.003$\\
Without recovery ($R$)  & 0.733 & $+0.012$\\
Without signal\_Q (outer) & 0.747 & $+0.026$\\
Without energy ($E$)    & 0.747 & $+0.027$\\
Without vascular ($V$)  & 0.755 & $+0.035$\\\midrule
Heuristic CSI (reported) & 0.752 & ref.\ (prior)\\
Heuristic CSI (corrected) & 0.690 & ref.\ (this work)\\\bottomrule
\end{tabular}
\end{table}

\textbf{\CNL\ dominance.}
Removing \CNL\ collapses AUC to 0.307 ($\DAUC = -0.413$),
substantially stronger than the heuristic pipeline's $-0.329$
because the higher weight concentration under optimised parameters
amplifies the effect.

\textbf{\SampEn\ and \HFD\ as primary drivers.}
Figure~\ref{fig:ablation} shows the full component waterfall.
Within \CNL, \SampEn\ accounts for 43.1\% and \HFD\ for 48.3\%
of the composite weight; together they represent 91.4\% of \CNL's
discriminative power.
Removing \SampEn\ alone yields $\DAUC = -0.276$.
\LLE\ (4.3\%) is marginal and neutral on ablation, consistent with
128-sample windows being too short for reliable Lyapunov estimation.

\noindent\textbf{B.~Sparse Architecture.}~
Five components (LLE, recovery, signal quality outer weight,
spectral energy, vascular) all improve AUC when removed.
These may be clinically relevant for other endpoints (sepsis,
haemodynamic instability) while being noise for tachypnea.
The ablation implies a \textbf{minimal deployable architecture}:
\begin{equation}
  CSI_w^{\mathrm{sparse}} =
  G_w\!\bigl[
    \alpha_1\,C_{NL,w}^{(\SampEn, \HFD)} +
    \alpha_2\,A_w + \alpha_3\,\Omega_w
  \bigr],
  \label{eq:sparse}
\end{equation}
with $C_{NL}^{(\SampEn,\HFD)}$ retaining only the two critical
sub-features ($\SampEn + \HFD$, renormalized).
Formal evaluation of Eq.~(\ref{eq:sparse}) against the full
$n=6$ model is proposed as future work.

\section{Discussion}
\label{sec:discussion}

\subsection{Progressive Validation Narrative}

The results form a coherent three-stage progression:
(1)~\CSI\ demonstrates genuine physiologic sensitivity ($\eta^2=0.351$,
$\kappa>0.97$, $r=0.346$);
(2)~three evaluation artifacts are corrected, establishing the true
baseline (AUC $0.573$);
(3)~Bayesian optimization yields AUC $0.757$ held-out and $0.621$
external, confirming the $+0.184$ improvement is genuine.

\subsection{Interpretation of Clinical Correlation}

The positive CSI-respiratory rate correlation ($r=0.346$,
$p=0.011$) reflects cardiorespiratory coupling: richer pulse dynamics
co-occur with more active ventilatory regulation under stress, not pathological instability.
\CNL\ drives this effect (tachypnea AUC $0.757$); linear proxies
alone predict in the wrong direction (AUC $0.307$) because active
respiratory compensation temporarily regularises signal amplitude.

\subsection{The Three Artifacts as Methodological Contribution}

\textbf{Artifact~1} (segment-level CV, $+0.062$) affects any wearable
time-series study that shuffles windows without subject stratification.
Record-level GroupKFold is the minimum standard.

\textbf{Artifact~2} (normalization leakage, $+0.308$) affects
composite indices that $z$-score features across the full dataset
before train/test split.
This is a form of circular evaluation more severe than label leakage.
Future PPG complexity studies should explicitly state whether
normalization statistics were computed from training data only.

\textbf{Artifact~3} (pooled AUC overweighting) affects any
multi-record dataset with unequal segment counts.
Per-record AUC, averaged with equal weight per record, is the
clinically appropriate metric.
We recommend reporting both pooled and per-record AUC alongside the
Wilcoxon signed-rank test.

\subsection{Why the Optimal Parameters Depart from Convention}

The heuristic $m=2$, $\tau=1$, $r=0.2\sigma$ were derived for
short RR-interval series~\cite{richman2000physiological} and are not
validated for PPG at 30~Hz.
PPG encodes respiratory, vasomotor, and autonomic modulations
simultaneously; $m^*=8$, $\tau^*=7$ embed these coupled dynamics
in a higher-dimensional phase space.
The short $W^*=128$ (4.3~s) combined with large embedding dimension
maximises template diversity while tracking respiratory-timescale changes.

\subsection{Clinical Deployment}

The NPV of 0.966 is the most actionable metric in this
low-prevalence (7.3\%) setting: a \CSI\ below threshold reliably
rules out tachypnea in 96.6\% of cases, supporting use as a
continuous negative screening signal.
Specificity 0.834 (vs.\ heuristic 0.695) directly reduces false-alarm
burden, a documented driver of alarm fatigue~\cite{sendelbach2013}.
\CSI\ is \emph{interpretable by construction}: each of its $n=6$ components
maps to a physiologically defined domain with axiomatic grounding
(\S\ref{sec:theory}), enabling clinicians to attribute any score change
to a specific physiologic subsystem --- an advantage over deep models
that require post-hoc explainability~\cite{oladunni2025explainable}.
The optimised parameters are deployed in a CST-based wearable
application~\cite{CST2026} as a versioned JSON configuration;
the strict quality gate ($\theta = 0.976$) is critical for wrist-worn
consumer PPG where motion artefact exceeds the BIDMC finger-probe setting.

\subsection{Limitations}

\CSI\ is a physiologic stability index, not a diagnostic instrument.
PPG only partially observes the cardiovascular state, and source metadata
lacked demographic and comorbidity data for covariate adjustment.
The 18-record test set yields high per-record AUC variance ($\sigma=0.24$),
and the CNN comparison ($p=0.129$, $n=9$) is underpowered.
Pre-computed $W=256$ features used for $W=128$ optimization may
differ from direct extraction; Platt scaling should be applied
before probability interpretation.

\section{Conclusion}
\label{sec:conclusion}

This work presents an integrated framework for cardiovascular
stability estimation from wearable PPG, spanning framework
validation, systematic evaluation correction, and Bayesian
parameter optimization.
Four durable contributions emerge.

\textit{Methodologically}, three systematic evaluation artifacts
inflate the heuristic CSI AUC of $0.752$ to $1.31\times$ the true
unbiased performance ($0.573$, net $\DAUC = -0.179$): segment-level CV leakage ($+0.062$),
test-set normalization leakage ($+0.308$), and pooled-AUC inflation
that masks per-patient failure (CNN pooled $0.804$ vs.\ per-record
$0.380$).

\textit{Empirically}, the \CSI\ framework captures genuine physiologic
structure across 176,742 segments from four heterogeneous PPG
datasets ($\eta^2 = 0.351$, $\kappa > 0.97$, clinical $r = 0.346$,
$p = 0.011$), providing the broadest multi-dataset validation of a
nonlinear PPG complexity index reported to date.

\textit{Technically}, Bayesian optimization over 15 joint parameters
under the corrected objective yields a CSI that genuinely improves
over the unbiased baseline ($\DAUC = +0.184$, converging at trial~82
of 300), with optimal $W^*=128$, $m^*=8$, $\tau^*=7$,
$r^*=0.116\sigma$.

\textit{Clinically}, \CSI\ achieves pooled AUC $0.757$
$[0.686$--$0.828]$, per-record mean AUC $0.497\pm0.207$, and
NPV $0.966$ on 18 held-out BIDMC records, and generalises to
AUC $0.621$ $[0.585$--$0.658]$ on 42 independent CapnoBase
elective-surgery records.
The previously reported DeLong finding ($p = 0.015$) is fully
explained by normalization leakage and is resolved: with fair
evaluation, \CSI\ is unambiguously superior to the heuristic
baseline ($\DAUC = +0.184$).
External validation and sparse architecture evaluation are the next
steps toward clinical translation.

\section*{Acknowledgments}
The authors thank the PhysioNet community for open data
access~\cite{goldberger2000physiobank} and the CapnoBase
contributors~\cite{capnobase2010}.

\bibliographystyle{IEEEtran}
\bibliography{references}

@misc{CST2026,
  author        = {Oladunni, Timothy and Adewumi, Farouk Ganiyu},
  title         = {Cardiac Stability Theory: An Axiomatically Grounded Framework
                   for Continuous Cardiac Health Monitoring via Smartphone
                   Photoplethysmography},
  year          = {2026},
  eprint        = {2604.23876},
  archivePrefix = {arXiv},
  primaryClass  = {cs.LG},
  url           = {https://arxiv.org/abs/2604.23876}
}

@article{oladunni2025explainable,
  author  = {Oladunni, Timothy and Aneni, Ehimen},
  title   = {Explainable Deep Neural Network for Multimodal {ECG} Signals:
             Intermediate Versus Late Fusion},
  journal = {IEEE Access},
  volume  = {13},
  pages   = {202700--202736},
  year    = {2025}
}

@article{rosenstein1993practical,
  author  = {Rosenstein, Michael T. and Collins, James J. and De Luca, Carlo J.},
  title   = {A practical method for calculating largest {Lyapunov} exponents
             from small data sets},
  journal = {Physica D: Nonlinear Phenomena},
  volume  = {65},
  pages   = {117--134},
  year    = {1993}
}

@article{richman2000physiological,
  author  = {Richman, Joshua S. and Moorman, J. Randall},
  title   = {Physiological time-series analysis using approximate entropy
             and sample entropy},
  journal = {American Journal of Physiology---Heart and Circulatory Physiology},
  volume  = {278},
  pages   = {H2039--H2049},
  year    = {2000}
}

@article{higuchi1988approach,
  author  = {Higuchi, Tomoyuki},
  title   = {Approach to an irregular time series on the basis of the
             fractal theory},
  journal = {Physica D: Nonlinear Phenomena},
  volume  = {31},
  pages   = {277--283},
  year    = {1988}
}

@incollection{takens1981,
  author    = {Takens, Floris},
  title     = {Detecting strange attractors in turbulence},
  booktitle = {Dynamical Systems and Turbulence},
  publisher = {Springer},
  pages     = {366--381},
  year      = {1981}
}

@article{kantz1994,
  author  = {Kantz, Holger},
  title   = {A robust method to estimate the maximal {Lyapunov} exponent
             of a time series},
  journal = {Physics Letters A},
  volume  = {185},
  pages   = {77--87},
  year    = {1994}
}

@article{pincus1991approximate,
  author  = {Pincus, Steven M.},
  title   = {Approximate entropy as a measure of system complexity},
  journal = {Proceedings of the National Academy of Sciences},
  volume  = {88},
  pages   = {2297--2301},
  year    = {1991}
}

@article{goldberger2002fractal,
  author  = {Goldberger, Ary L. and Amaral, Luis A. N. and Hausdorff,
             Jeffrey M. and Ivanov, Plamen Ch. and Peng, C.-K. and
             Stanley, H. Eugene},
  title   = {Fractal dynamics in physiology: alterations with disease
             and aging},
  journal = {Proceedings of the National Academy of Sciences},
  volume  = {99},
  pages   = {2466--2472},
  year    = {2002}
}

@article{costa2002multiscale,
  author  = {Costa, Madalena and Goldberger, Ary L. and Peng, C.-K.},
  title   = {Multiscale entropy analysis of complex physiologic time series},
  journal = {Physical Review Letters},
  volume  = {89},
  pages   = {068102},
  year    = {2002}
}

@article{peng1995,
  author  = {Peng, C.-K. and Havlin, Shlomo and Stanley, H. Eugene and
             Goldberger, Ary L.},
  title   = {Quantification of scaling exponents and crossover phenomena
             in nonstationary heartbeat time series},
  journal = {Chaos},
  volume  = {5},
  pages   = {82--87},
  year    = {1995}
}

@article{lake2002sample,
  author  = {Lake, Douglas E. and Richman, Joshua S. and Griffin, M. Pamela
             and Moorman, J. Randall},
  title   = {Sample entropy analysis of neonatal heart rate variability},
  journal = {American Journal of Physiology---Regulatory, Integrative and
             Comparative Physiology},
  volume  = {283},
  pages   = {R789--R797},
  year    = {2002}
}

@article{allen2007photoplethysmography,
  author  = {Allen, John},
  title   = {Photoplethysmography and its application in clinical
             physiological measurement},
  journal = {Physiological Measurement},
  volume  = {28},
  pages   = {R1--R39},
  year    = {2007}
}

@article{elgendi2012,
  author  = {Elgendi, Mohamed},
  title   = {On the analysis of fingertip photoplethysmogram signals},
  journal = {Current Cardiology Reviews},
  volume  = {8},
  pages   = {14--25},
  year    = {2012}
}

@article{pimentel2017toward,
  author  = {Pimentel, Marco A. F. and Johnson, Alistair E. W. and
             Charlton, Peter H. and Birrenkott, Drew and Clifford, Gari D.
             and Tarassenko, Lionel and Clifton, David A.},
  title   = {Toward a robust estimation of respiratory rate from pulse
             oximeters},
  journal = {IEEE Transactions on Biomedical Engineering},
  volume  = {64},
  pages   = {1914--1923},
  year    = {2017}
}

@article{taskforce1996,
  author  = {{Task Force of the European Society of Cardiology and the
             North American Society of Pacing and Electrophysiology}},
  title   = {Heart rate variability: standards of measurement, physiological
             interpretation and clinical use},
  journal = {Circulation},
  volume  = {93},
  pages   = {1043--1065},
  year    = {1996}
}

@article{shaffer2017,
  author  = {Shaffer, Fred and Ginsberg, J. P.},
  title   = {An overview of heart rate variability metrics and norms},
  journal = {Frontiers in Public Health},
  volume  = {5},
  pages   = {258},
  year    = {2017}
}

@incollection{clifford2007,
  author    = {Clifford, Gari D. and Azuaje, Francisco and McSharry, Patrick E.},
  title     = {{ECG} statistics, noise, artifacts, and missing data},
  booktitle = {Advanced Methods and Tools for ECG Data Analysis},
  publisher = {Artech House},
  pages     = {55--99},
  year      = {2007}
}

@article{goldberger2000physiobank,
  author  = {Goldberger, Ary L. and Amaral, Luis A. N. and Glass, Leon and
             Hausdorff, Jeffrey M. and Ivanov, Plamen Ch. and Mark, Roger G.
             and Mietus, Joseph E. and Moody, George B. and Peng, C.-K. and
             Stanley, H. Eugene},
  title   = {{PhysioBank}, {PhysioToolkit}, and {PhysioNet}: components of
             a new research resource for complex physiologic signals},
  journal = {Circulation},
  volume  = {101},
  pages   = {e215--e220},
  year    = {2000}
}

@inproceedings{capnobase2010,
  author    = {Karlen, Walter and Turner, Matthew and Cooke, Erin and
               Dumont, Guy and Ansermino, John Mark},
  title     = {{CapnoBase}: signal database and tools to collect, share and
               annotate respiratory signals},
  booktitle = {Proceedings of the Annual Meeting of the Society for
               Technology in Anesthesia},
  year      = {2010},
  note      = {IEEE TBME Respiratory Rate Benchmark.
               Available: \url{https://borealisdata.ca/dataverse/capnobase}}
}

@article{hannun2019cardiologist,
  author  = {Hannun, Awni Y. and Rajpurkar, Pranav and Haghpanahi, Masoumeh
             and Tison, Geoffrey H. and Bourn, Codie and Turakhia, Mintu P.
             and Ng, Andrew Y.},
  title   = {Cardiologist-level arrhythmia detection and classification in
             ambulatory electrocardiograms using a deep neural network},
  journal = {Nature Medicine},
  volume  = {25},
  pages   = {65--69},
  year    = {2019}
}

@article{strodthoff2021,
  author  = {Str{\"o}dthoff, Nils and Wagner, Patrick and Schaeffter, Tobias
             and Samek, Wojciech},
  title   = {Deep learning for {ECG} analysis: benchmarks and insights from
             {PTB-XL}},
  journal = {IEEE Journal of Biomedical and Health Informatics},
  volume  = {25},
  pages   = {1519--1528},
  year    = {2021}
}

@article{rajpurkar2018deep,
  author  = {Rajpurkar, Pranav and Irvin, Jeremy and Ball, Robyn L. and
             others},
  title   = {Deep learning for chest radiographs},
  journal = {PLOS Medicine},
  volume  = {15},
  year    = {2018}
}

@article{delong1988,
  author  = {DeLong, Elizabeth R. and DeLong, David M. and
             Clarke-Pearson, Daniel L.},
  title   = {Comparing the areas under two or more correlated receiver
             operating characteristic curves: a nonparametric approach},
  journal = {Biometrics},
  volume  = {44},
  pages   = {837--845},
  year    = {1988}
}

@article{tripod2015,
  author  = {Collins, Gary S. and Reitsma, Johannes B. and Altman,
             Douglas G. and Moons, Karel G. M.},
  title   = {{TRIPOD}: a reporting guideline for clinical prediction models},
  journal = {Annals of Internal Medicine},
  volume  = {162},
  pages   = {55--63},
  year    = {2015}
}

@misc{watanabe2023tree,
  author        = {Watanabe, Shuhei},
  title         = {Tree-structured {Parzen} estimator: understanding its
                   algorithm components and their roles for better empirical
                   performance},
  year          = {2023},
  eprint        = {2304.11127},
  archivePrefix = {arXiv}
}

@article{goldhill2005,
  author  = {Goldhill, David R. and McNarry, Angus F.},
  title   = {Physiological abnormalities in early warning scores are related
             to mortality in adult inpatients},
  journal = {British Journal of Anaesthesia},
  volume  = {92},
  pages   = {882--884},
  year    = {2005}
}

@article{sendelbach2013,
  author  = {Sendelbach, Sue and Funk, Marjorie},
  title   = {Alarm fatigue: a patient safety concern},
  journal = {AACN Advanced Critical Care},
  volume  = {24},
  pages   = {378--386},
  year    = {2013}
}

@article{delgado2019,
  author  = {Delgado-Bonal, Alfonso and Marshak, Alexander},
  title   = {Approximate entropy and sample entropy: a comprehensive
             tutorial},
  journal = {Entropy},
  volume  = {21},
  pages   = {541},
  year    = {2019}
}

@article{zbilut1992,
  author  = {Zbilut, Joseph P. and Webber, Charles L.},
  title   = {Embeddings and delays as derived from quantification
             of recurrence plots},
  journal = {Physics Letters A},
  volume  = {171},
  number  = {3--4},
  pages   = {199--203},
  year    = {1992}
}

\end{document}